\documentclass{article}

\usepackage{graphicx}
\usepackage{authblk}
\usepackage[caption=false]{subfig}
\usepackage{amsthm}
\usepackage{amsmath}
\usepackage{amssymb}
\usepackage[margin=1.25in]{geometry}
\usepackage{color}
\usepackage{morefloats}
\usepackage{epstopdf}

\theoremstyle{definition}

\usepackage{hyperref}

\begin{document}

\title{The Formation and Imprinting of Network Effects Among the Business Elite}

\author[1,2]{Brian Uzzi}
\author[1]{Yang Yang}
\author[2]{Kevin Gaughan}
\affil[1]{Northwestern Institute on Complex Systems (NICO)\\
Northwestern University}
\affil[2]{Kellogg School of Management\\
Northwestern University}

\maketitle

\begin{abstract}
The ``business elite'' constitutes a small but strikingly influential subset of the population, oftentimes affecting important societal outcomes such as the consolidation of political power \cite{padgett:1993}, the adoption of corporate governance practices, and the stability of national economies more broadly. Research has shown that this exclusive community often resembles a densely structured network, where elites exchange privileged access to capital, market information, and political clout in an attempt to preserve their economic interests and maintain the status quo \cite{useem:1982}. While there is general awareness that connections among the business elite arise because ``elites attend the same schools, belong to the same clubs, and in general are in the same place at the same time'', surprisingly little is known about the network dynamics that emerge within these formative settings. Here we analyze a unique dataset of all MBA students at a top 5 MBA program. Students were randomly assigned to their first classes; friendship among students prior to coming into the program was rare; and the network data – email transmissions among students – were collected for the year 2006 when students almost entirely used the school's email server to communicate, thereby providing an excellent proxy for their networks. After matching students on all available characteristics (e.g., age, grade scores, industry experience, etc.) — i.e. creating ``twin pairs'' — we find that the distinguishing characteristics between students who do well in job placement and those who do not is their network. Further, we find that the network differences between the successful and unsuccessful students develops within the first month of class and persists thereafter, suggesting a network imprinting that is persistent. Finally, we find that these effects are pronounced for students who are at the extreme ends of the distribution on other measures of success – students with the best expected job placement do particularly poorly without the right network (``descenders''), whereas students with worst expected job placement pull themselves to the top of the placement hierarchy (``ascenders'') with the right network.
\end{abstract}

\section*{Data}
We collected and analyzed more than 4.5 million time-stamped emails from students at a globally top-ranked MBA program, focusing specifically on the relationship between students' evolving communication networks and their subsequent career outcomes. This data is available in the form of email logs recorded and stored by the university, along with registrar data on each student before and during their matriculation in the program. Included in the dataset is a record of each email sent by an MBA student between Fall 2006 and Spring 2008. The record includes the date and time at which the email was sent and received and the (anonymous) numeric IDs of the sender and receiver of the message. Academic records (GMAT scores, grades, extra-curricular activities, prior work experience and job titles), and demographic data (age, race and nationality), were merged with the network data to connect email transmissions with personal characteristics. There are approximately 11.5 million e-mails and ~4.5 million student-student e-mails in the data. An important characteristics of the data is its randomized design.  Students are randomly assigned to sections within the school, minimizing selection effects. Since observations began when students first met each other, we eliminated the left censoring that typically occurs when network data are captured after ties have already been formed.

\begin{figure}[ht]
	\centerline{\includegraphics[height=4in]{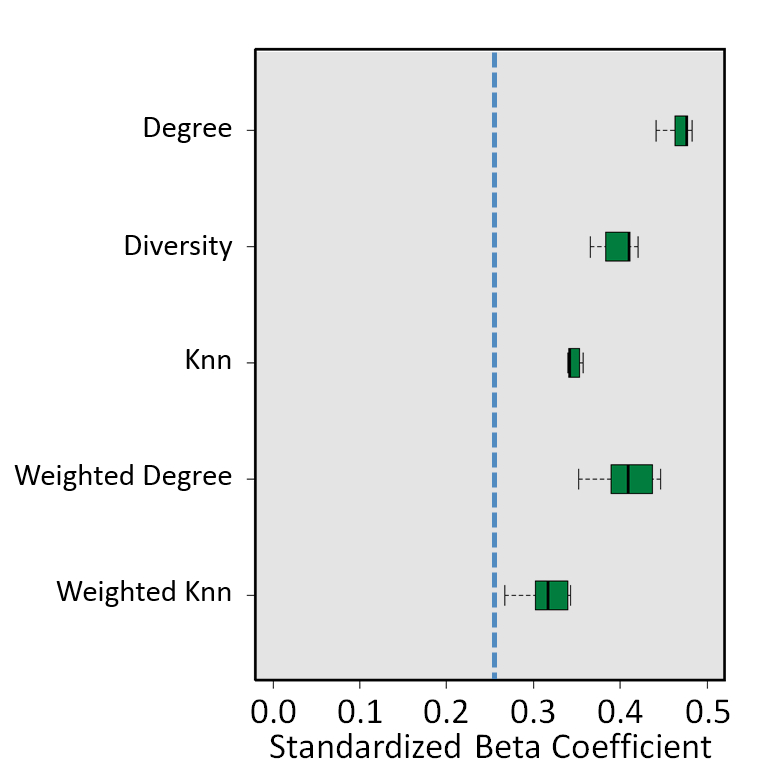}}
	\caption{The effect of network structure on post-graduation salary rank. This figure reports the standardized beta coefficients of OLS regression models using coarsened exact matching \cite{king:2011}. In all models, the dependent variable is the relative post-graduation salary rank compared to one's peers. Students were matched identically along the following covariates: gender, ethnicity, industry, years of experience, and GPA. Twelve beta coefficients were estimated for each variable-one for each month that a student was enrolled in the program. The box-plots for each network measure indicate a positive association with salary rank. Additionally, the short length of the box-plot whiskers indicates that there was little variance in the measures from month to month.}
\label{fig_1}
\end{figure}

\section*{Methods}
In order to understand the causal relationship between network effects and job rank, we used coarsened exact matching (CEM) (see \cite{king:2011}) to construct a reduced, matched sample (matching students on all characteristics possible, i.e., age, GPA, industry experience). Then on the matched data, we further examine the significance of network effects on students' job rank. Additionally, in order to demonstrate that our observations cannot be explained by a random network process, we compare real observations to the null model where the degree sequence of network is preserved and links are placed completely randomly.

\section*{Results}
\begin{figure}[h]
	\centerline{\includegraphics[height=2.5in]{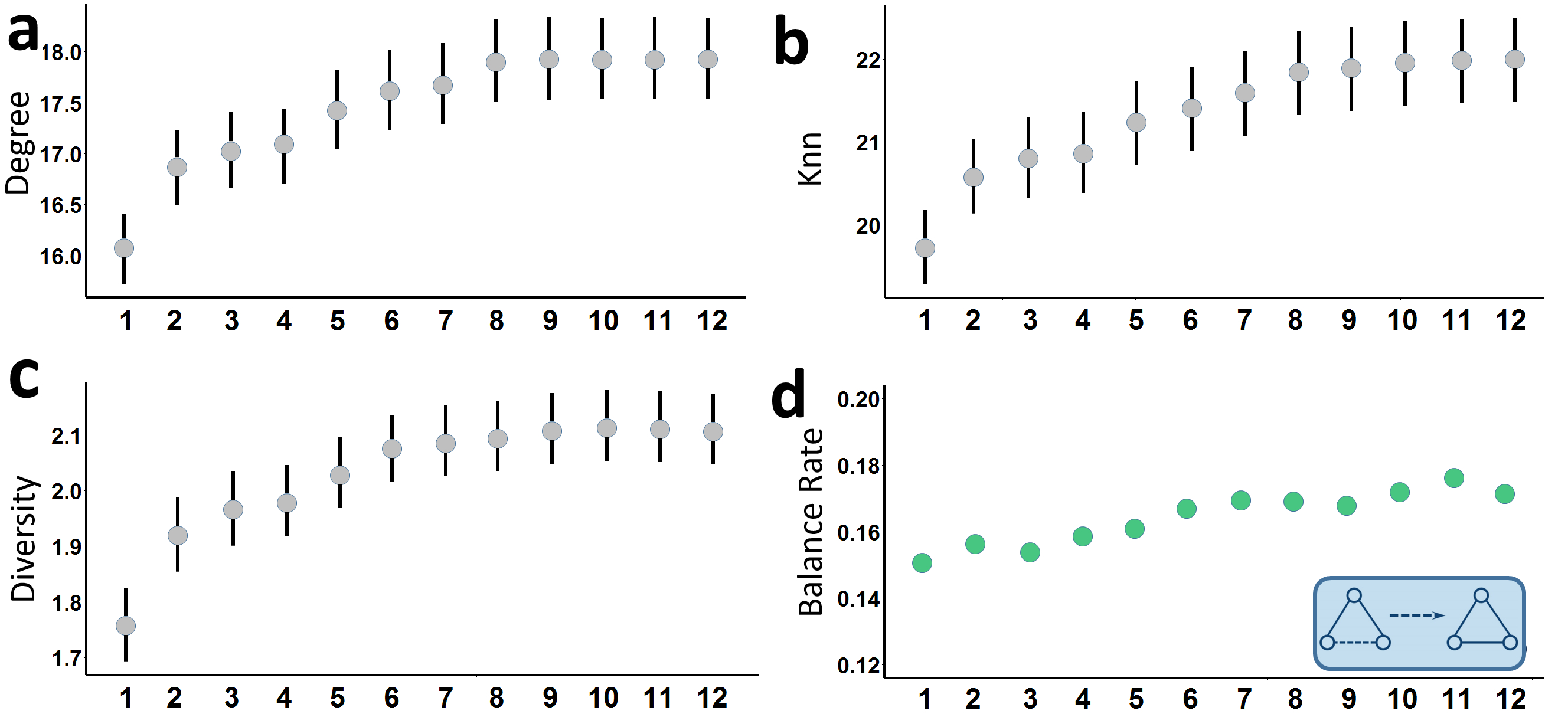}}
	\caption{Time-series of network properties among MBA students. a) A node's average number ties (degrees) over time. b) The average nearest neighbor degree (Knn), which represents the average degree of a node's neighbors. c) The structural balance rate of a student's network, measuring the effect of triadic closure in network. Note that in all sub-figures, there is a sharp increase between month 0 and month 1, which diminishes at an exponential rate during the remainder of time in the program.}
	\label{fig_2}
\end{figure}

\begin{figure}[h]
	\centerline{\includegraphics[height=4in]{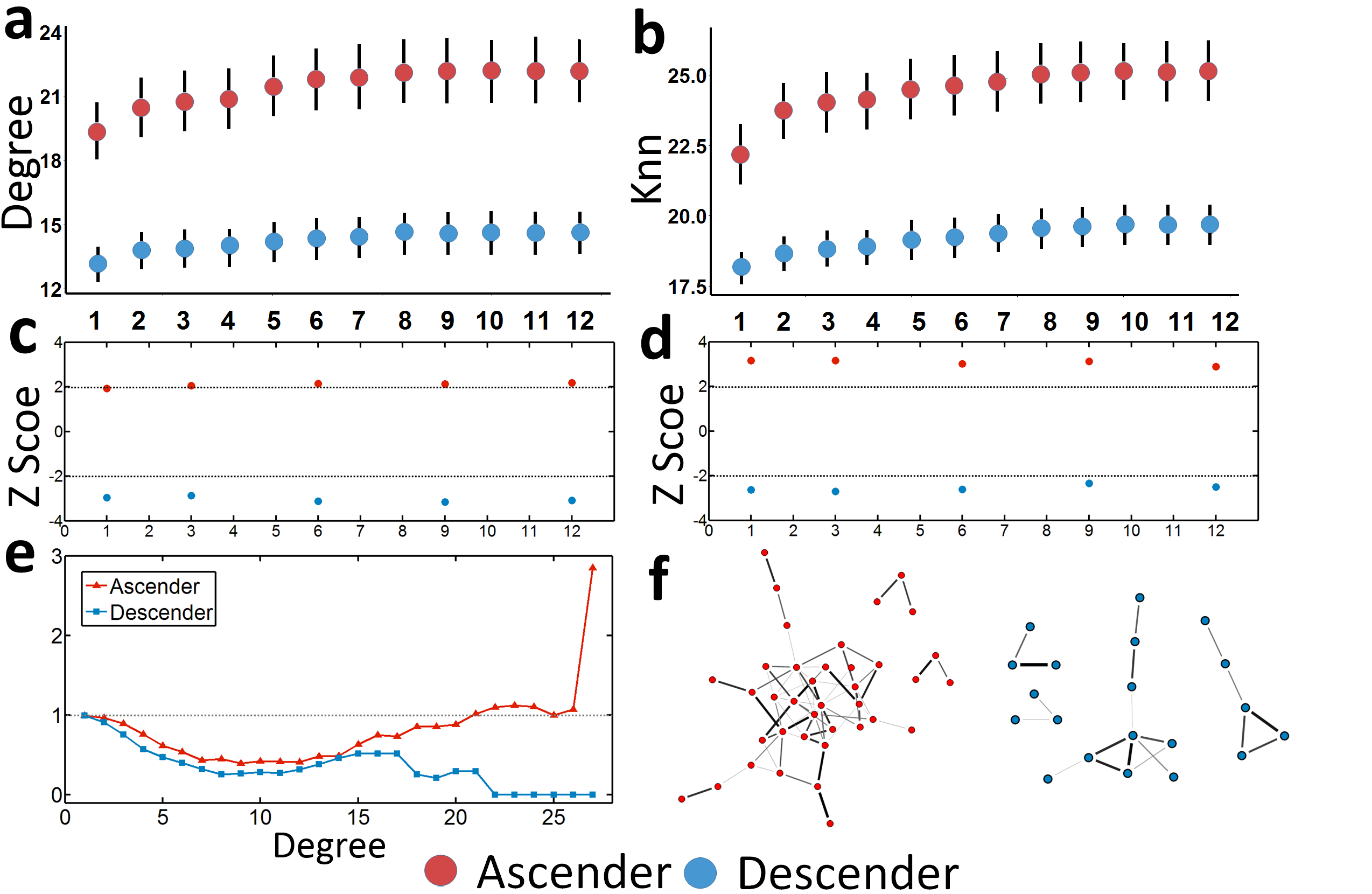}}
	\caption{Differences in network patterns among ascenders and descenders. In this figure, ascenders (descender) are defined as students who enter the program with a GMAT score in the lowest (highest) quartile and graduate with a salary rank in the top (bottom) quartile. a-b) The degree (Knn, average nearest neighbor degree) time series of ascenders and descenders over 12 months with 95\% confidence interval. c-d) This figure compares the degrees (Knn) between ascenders and descenders to the random network where the types of students are randomly assigned. e) The normalized rich-club coefficients p(k) for ascenders and descenders, which demonstrates a clear bifurcation for students with degree greater than 15. f) The sub-networks of ascenders and descenders with degrees larger than 10. This provides a more straightforward view that ascenders are more densely connected while descenders are sparsely connected.}
	\label{fig_3}
\end{figure}

Three important findings emerge from the data. First, a student's likelihood of securing a coveted, high paying job after graduation is strongly associated with the type of network they develop during their time in the program. Students with a higher network degree, greater network centrality, and more balanced communication among alters tend to have the highest post-graduation salaries, even when matching students along several covariates (Figure~\ref{fig_1}). Second, and somewhat surprisingly, we find that the structural characteristics of a student's ego network emerge as early as one month into the program and remain remarkably stable thereafter (Figures~\ref{fig_1}~\ref{fig_2}). The above findings suggest that, when exposed to a new social system, actors' initial network configurations may provide ``early warning signals of success'' — a finding that has important practical and policy-level implications. Lastly, we observe robust patterns of ``rich-club'' behavior at the level of the global network, where highly central students are more densely connected among themselves than students of a lower degree \cite{colizza:2006} \cite{yang:2014}. While the students who comprise the rich-club networks were more likely to have a higher salary post-graduation, they were also less likely to have the highest rank on other objective indicators of ability, such as GMAT scores and GPA (Figure~\ref{fig_3}). We interpret these findings as evidence of a trade-off between the development of human capital and social capital during an MBA program: On the one hand, students can choose to invest in building technical skills and domain-specific knowledge to enhance their career prospects; on the other hand, they can choose to invest in building their social capital by developing new ties and fostering a robust community of peers. Though these choices are certainly not mutually exclusive, the economic benefits of the former appear to significantly outweigh the benefits of the latter.

\section*{Conclusion}
Despite the fact that business schools advertise their role in fostering a valuable, life-long network (Haas MBA 2013), we find considerable differences in the types of networks that students actually develop - differences that are strongly linked to their future job placement and, ultimately, their access to the inner circles of the managerial elite. 

Our findings have important implications for both MBA students and the firms that hire them. From the standpoint of a student, the data suggest that important resources exist within an emerging MBA network that ultimately improve one's attractiveness to employers. While we can only speculate about the mechanism underling this pattern of results — i.e. specific networks may foster the development of particular social skills or, instead, they may provide access to employer networks outside of the program itself — we nevertheless find that it quite literally ``pays'' to develop one's social network in the early stages of a program. From the standpoint of prospective recruiters, the subtle signals that allow one to reliably infer a student's network structure \cite{yang:2012} may provide valuable insight into the qualities that the applicant will bring with them to their new job.


\begin{thebibliography}{10}
\bibitem{padgett:1993}
John F. Padgett, Christopher K. Ansell. American Journal of Sociology, Volume 98, Issue 6 (May, 1993), 1259-1319.

\bibitem{useem:1982}
Michael Useem. Classwide Rationality in the Politics of Managers and Directors of Large Corporations in the United States and Great Britain. Administrative Science Quarterly Vol. 27, No. 2 (Jun., 1982), pp. 199-226

\bibitem{king:2011}
Stefano M Iacus, Gary King, and Giuseppe Porro. 2011. Multivariate Matching Methods That are Monotonic Imbalance Bounding. Journal of the American Statistical Association, 493, 106: 345-361, 2011.

\bibitem{colizza:2006}
Vittoria Colizza, Alessandro Flammini, M. Angeles Serrano, and Alessandro Vespignani. Detecting rich-club ordering in complex networks. Nature Physics 2, 110-115 (2006), doi:10.1038/nphys209.

\bibitem{yang:2012}
Yang Y., Chawla, N.~V., Sun, Y., and Han, J. Link Prediction in Heterogeneous Networks: Influence and Time Matters. Proc. of the 12th IEEE International Conference on Data Mining (ICDM'12), 2012.

\bibitem{yang:2014}
Yang Y., Chawla, N.~V., and Dong, Y. Predicting Node Degree Centrality with the Node Prominence Profile. Nature Scientific Reports, doi:10.1038/srep07236, 2014.

\end{thebibliography}
\end{document}